\documentclass[a4paper,10pt,twoside]{cpc-hepnp}
\usepackage{upgreek,fancyhdr}
\usepackage{multicol,color}
\usepackage{graphicx,subfigure,dcolumn}
\usepackage{epsfig}
\usepackage{booktabs,slashed}
\usepackage{amssymb,bm,mathrsfs,bbm,amscd,amsfonts}
\usepackage[tbtags]{amsmath}
\usepackage{lastpage}
\usepackage{multirow}

\newcommand{\f}{\begin{equation}}
\newcommand{\ff}{\end{equation}}
\newcommand{\fa}{\begin{eqnarray}}
\newcommand{\ffa}{\end{eqnarray}}

\providecommand{\U}[1]{\protect\rule{.1in}{.1in}}
\begin{document}

\title{\boldmath The instability of $AdS$ black holes with lattices
\thanks{We are very grateful to Peng Liu, Yuxuan Liu, Chao Niu, Jianpin Wu
and Zhuoyu Xian for helpful discussions and former collaborations
on building holographic lattices. This work is supported by the
Natural Science Foundation of China under Grant No. 11875053.}}

\author{Yi Ling  $^{1,2,3}$ \email{lingy@ihep.ac.cn}
\quad Meng-He Wu  $^{1,2}$ \email{mhwu@ihep.ac.cn}}

\maketitle

\address{
$^1$ Institute of High Energy Physics, Chinese Academy of Sciences, Beijing, 100049, China \\
$^2$ School of Physics, University of Chinese Academy of Sciences, Beijing, 100049, China \\
$^3$ Shanghai Key Laboratory of High Temperature Superconductors, Shanghai,200444, China}
\maketitle

\begin{abstract}
The Anti-de Sitter (AdS) black hole with lattice structure plays
an essential role in the study of the optical conductivity in
holographic approach. We investigate the instability of this sort
of black holes which may lead to the holographic description of
charge density waves. In the presence of homogeneous axion fields,
we show that the instability of AdS-Reissner-Nordstr\"om(AdS-RN) black hole is always
suppressed. However, in the presence of Q-lattices, we find that
the unstable region becomes the smallest in the vicinity of
the critical region for metal/insulator phase transition. This
novel phenomenon is reminiscent of the behavior of the holographic
entanglement entropy during quantum phase transition.

\end{abstract}

\begin{keyword}
gauge/gravity duality, holographic gravity
\end{keyword}

\begin{pacs}
11.25.Tq, 04.70.bw
\end{pacs}


\section{Introduction}

In recent years the gauge/gravity duality has  successfully been
applied to the strongly coupled system in condense matter physics,
 which now as an important branch of holographic approach is dubbed as AdS/CMT duality \cite{Hartnoll:2016apf}.
It is well known that the instability of AdS black hole plays an
essential role in order to achieve the abundant phase structure of
the condensed matter system on the boundary. Especially, the
spontaneous breaking of $U(1)$ gauge symmetry in the bulk leads to a
new sort of black holes with scalar hair, which provides a novel
picture for the condensation of
superconductivity\cite{Gubser:2008px,Hartnoll:2008vx,Hartnoll:2008kx}.
While the spontaneous breaking of translational invariance leads
to the generation of spatially modulated modes for black holes,
which is holographically dual to the formation of charge density
waves
(CDW)\cite{Donos:2013gda,Ling:2014saa,Kiritsis:2015hoa,Jokela:2014dba,Cremonini:2016rbd,
Cremonini:2017usb,Cremonini:2018xgj,Andrade:2017leb,Jokela:2017ltu,Cremonini:2019fzz,
Amoretti:2019kuf,Song:2018uds}. For high $T_c$ superconductor, the
charge density wave phase is also named as the pseudo-gap phase,
which plays a key role in understanding the formation of high
$T_c$ superconductivity\cite{Ling:2019gjy}.

Originally the holographic CDW is formed over an AdS-RN background
with translational symmetry. The basic idea is to introduce
unstable terms into the action such that the $BF$ bound of AdS
black holes is violated below some value of the
Hawking temperature. As a result, the ordinary AdS-RN
background becomes unstable and flows to a new configuration which
exhibits a periodic structure along a spatial direction, giving
rise to the formation of CDW. In this paper we intend to
investigate the instability of a sort of AdS-RN black holes
{\it without} translational invariance. In AdS/CMT duality,
it is essential to construct a bulk geometry without translational
invariance in order to obtain a finite direct current (DC)
conductivity for the dual system. If the translational symmetry is
preserved, then the momentum would be conserved such that the DC
would flow without relaxation, giving rise to an infinite
conductivity. This of course is not the feature of practical
materials. Now in holographic framework, typically one has two
ways to break the translational symmetry by hand (for brief
review see \cite{Ling:2015ghh}). One way is to construct a
lattice manifestly by introducing spatially modulated sources in
the bulk
\cite{Horowitz:2012ky,Horowitz:2012gs,Horowitz:2013jaa,Ling:2013aya,Ling:2013nxa}.
However, in this framework it is very challenging to explore the
low temperature effects due to the numerical difficulties
involved in solving partial differential equations (PDEs)
\cite{Hartnoll:2014gaa}. An alternative way is to introduce the
momentum dissipation by linear axion fields, helical lattice or
Q-lattices, which might be called as ``homogeneous
lattices"\cite{Andrade:2013gsa,Donos:2012js,Donos:2013eha,Ling:2014laa,
Davison:2014lua,Baggioli:2014roa,Gouteraux:2016wxj,Andrade:2017cnc,
Amoretti:2017axe,Li:2018vrz,Ling:2015dma,Ling:2016wyr,Baggioli:2020nay}.
In these models, though the translation symmetry is broken, the
equations of motion are still ordinary differential equations
(ODEs) which can be numerically solved even in the zero
temperature limit. Remarkably, in this framework
 a novel metal/insulator transition has been observed\cite{Donos:2012js}, which makes it plausible to study
the quantum critical phenomenon in the holographic setup with
lattices\cite{Ling:2015dma}.

Therefore, it is very desirable to construct holographic CDW over
a lattice background rather than a background with translational
symmetry. This is the main motivation of this paper. As  the first
step, we will investigate the instability of a sort of
AdS black holes with homogeneous lattices by perturbative
analysis. We will determine the unstable region in the
configuration space in the presence of axion fields and
Q-lattices, respectively.  We will demonstrate that the presence
of  the linear axion field always suppresses the instability of
AdS-RN black hole. However, in Q-lattice framework, we find that
in low temperature limit, the unstable region becomes the smallest
near the critical region of the metal/insulator transition.
This novel phenomenon is reminiscent of the role of holographic
entanglement entropy, which signals the occurrence of quantum
phase transition.

This paper is organized as follows. In Section $2$ ,
we introduce the holographic model for charge density
waves, and briefly review the instability of the AdS-RN black
hole. In Section $3$,
we analyze the instability of
the black hole with momentum relaxation due to  the linear axion fields. Then we
focus on the instability of the Q-lattice background in Section
$4$.
Our results and conclusions are given in Section
$5$.

\section{The holographic setup} \label{sec1}
In this section, we introduce a holographic model without lattice
structure in four dimensional spacetime, in which the gravity is
coupled to a dilaton field and two massless $U(1)$ gauge fields.
Then in next sections we will impose lattice structure based on
this setup. The action is given by,
\begin{equation}\label{eq:eps+1}
  \begin{aligned}
  S=&\frac{1}{2\kappa^2}\int d^4 x \sqrt{-g} \ \mathfrak{L_1},
  \end{aligned}
\end{equation}
where
\begin{equation}\label{eq:eps+2}
  \begin{aligned}
  \mathfrak{L_1}&=& R-\frac{1}{2}\left ( \nabla \Phi  \right )^2-V\left ( \Phi  \right ) -\frac{1}{4}Z_{A}(\Phi )F^2  -\frac{1}{4} Z_{B}(\Phi )G^2 -\frac{1}{2} Z_{AB}(\Phi )FG,
  \end{aligned}
\end{equation}
with $F=dA$, $G=dB$. Two gauge fields $A$ and $B$ correspond to
two global $U(1)$ symmetries on the boundary. We will treat gauge
field $B$ as the electromagnetic field and consider its transport
properties. The real dilaton field $\Phi$ will be viewed as the
order parameter of translational symmetry breaking. We propose
that functions $V,\ Z_A, \ Z_{B}, \ Z_{AB}$ have the following
form:
\begin{equation}\label{eq:eps+3}
\begin{aligned}
V ( \Phi )&=-\frac{1}{L^{2}} + \frac{1}{2}m_{s}^{2}\Phi ^{2}, \\
 Z_{A}(\Phi )&=1-\frac{\beta }{2} L^2 \Phi^2,\\
 Z_{B}(\Phi )&=1,\\
 Z_{AB}(\Phi )&=\frac{\gamma }{\sqrt{2}} L \Phi.
\end{aligned}
\end{equation}

In above setup two important terms are introduced. One is the
$\beta$-term, which plays an essential role in inducing the
instability of AdS-RN  black holes to form CDW, while the other
$\gamma$-term is not essential, just driving the tip of the
unstable dome deviating from  $k_c= 0$.

The equations of motion are given by
\begin{equation}\label{eq:eps+4}
  \begin{aligned}
     & R_{\mu \nu }-T_{\mu \nu }^{\Phi }-T_{\mu \nu }^{A }-T_{\mu \nu }^{B }-T_{\mu \nu }^{AB  }  =0, \\
     & \nabla  ^{2} \Phi -\frac{1}{4}Z_{A}' F^2 -\frac{1}{4}Z_{B}' G^2 -V'=0,                                              \\
     & \nabla_{\mu }(Z_A F^{\mu \nu }+Z_{AB} G^{\mu \nu })=0,                                                                                   \\
     & \nabla_{\mu }(Z_B G^{\mu \nu} +Z_{AB} F^{\mu \nu } ) =0,
  \end{aligned}
\end{equation}
where
\begin{equation}\label{eq:eps+5}
  \begin{aligned}
    T_{\mu \nu }^{\Phi }   & =\frac{1}{2} \nabla_{\mu }\Phi \nabla_{\nu  }\Phi +\frac{1}{2}V g_{\mu \nu },                                       \\
    T_{\mu \nu }^{A  }     & =\frac{Z_{A}}{2} \left(F_{\mu \rho  }F^{\rho }_{\nu }- \frac{1}{4} g_{\mu \nu }F^2\right),                          \\
    T_{\mu \nu }^{B  }     & =\frac{Z_{B}}{2} \left(G_{\mu \rho  }G^{\rho }_{\nu }- \frac{1}{4} g_{\mu \nu }G^2\right),                          \\
    T_{\mu \nu }^{AB }  & = Z_{AB} \left(G_{ (\mu | \rho |  }F^{\rho }_{\nu )}- \frac{1}{4} g_{\mu \nu }GF\right).
  \end{aligned}
\end{equation}

The equations of motion admit the planar  AdS-RN black hole as a
solution with translational symmetry along both $x$ and $y$
directions, which is given as
\begin{equation}\label{eq:eps+6}
  \begin{aligned}
     & ds^{2}=\frac{1}{z^2}\left[-(1-z)p(z)dt^{2}+\frac{dz^{2}}{(1-z)p(z)}+dx^{2} +dy^{2}\right], \\
     & A_t=\mu(1-z), \ \Phi=0, \ B=0,
  \end{aligned}
\end{equation}
where $p(z)=4 \left(1+z+z^2-\frac{\mu^2  z^3}{16} \right)$, $\mu$
is the chemical potential of gauge field $A$. In this coordinate
system, the black hole horizon is located at $z=1$ and the AdS
boundary is at $z=0$. The Hawking temperature of the black hole is
simply given by $T/ \mu=(48-\mu^2)/(16 \pi \mu)$. Throughout this
paper we shall set the AdS radius $l^2=6L^2=1/4$. Without loss of generality,
two coupling constants will be set as $\beta=-94$ and $\gamma
=16.4$.
\begin{figure} [t]
  \center{
  \includegraphics[scale=0.45]{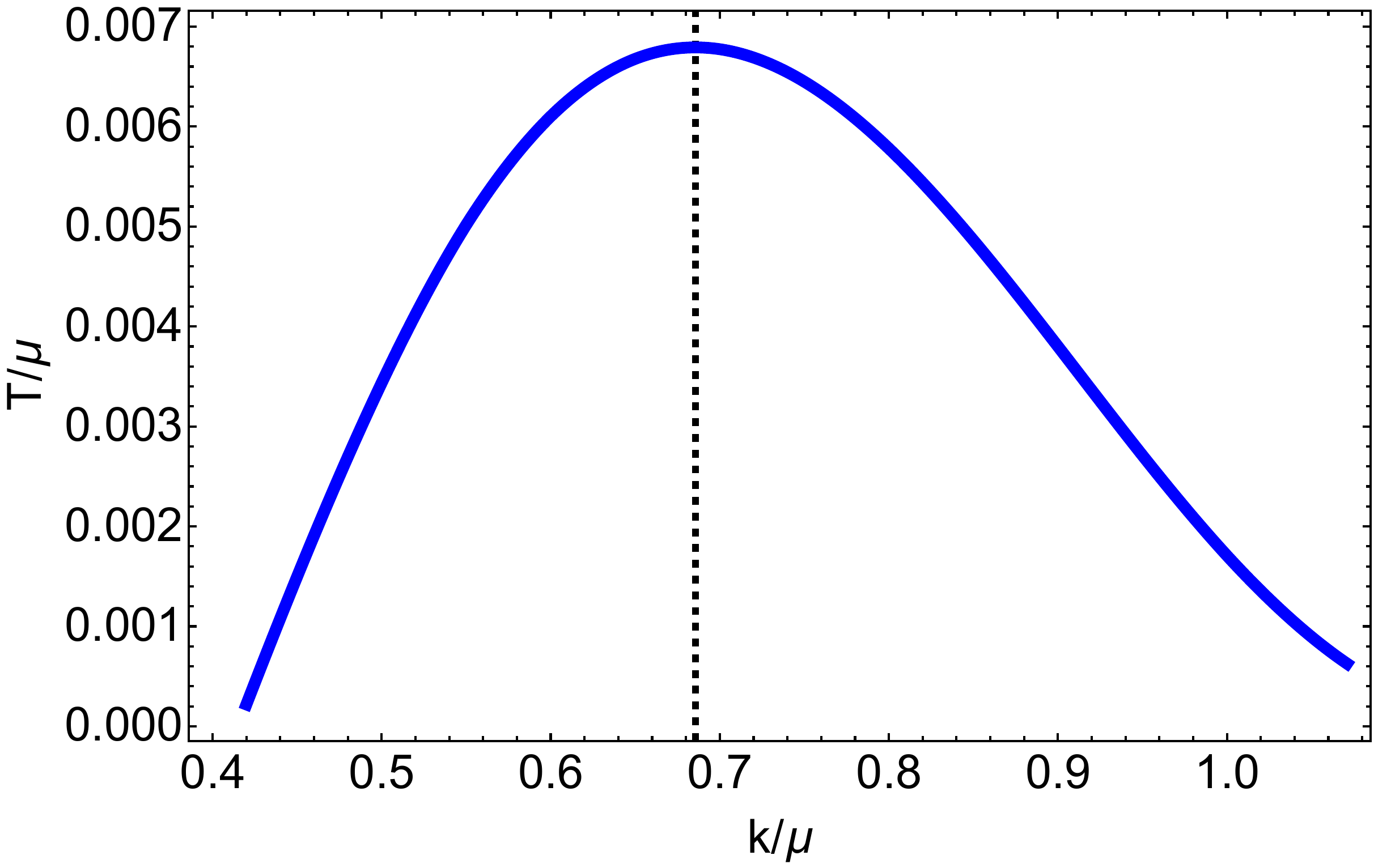}\ \hspace{0.1cm}
\caption{\label{fig1} The critical temperature as the
function of the wave number. Below the curve is the unstable
region which looks like a dome. The dashed line marks the location
of the tip of the dome, with $k_c \approx 0.6856$.
  }}
  \end{figure}

We remark that to analyze the instability of AdS black holes,
the key point is that in zero temperature limit, the near horizon
geometry of AdS-RN black holes is $AdS_2 \times R^2$. Once the $BF$
bound of $AdS_2$ is violated, the near horizon geometry will
become unstable and flow to a new solution as the infrared fixed
point, which is characterized by the appearance of the spatially
modulated modes in the bulk. And such analogous instability
remains at finite temperature for AdS-RN black holes (for details
we refer to \cite{Donos:2013gda}). Now specifically, we consider
the following perturbations to examine the instability of the
electrically charged AdS-RN black hole,
\begin{equation}\label{eq:eps+70}
  \begin{aligned}
  &\delta \Phi =\phi (z) \cos (k x), \\
 &\delta B =b_{t} (z) \cos (k x).
  \end{aligned}
\end{equation}
The instability of the background will be signaled by the
existence of non-trivial solutions to the perturbation equations
of $\delta \Phi$ and $\delta B$, which  spontaneously  break the
translational symmetry along $x$ direction. Taking the AdS-RN
as the background and substituting (\ref{eq:eps+70}) into the
equations of motion (\ref{eq:eps+4}), we obtain two coupled linear
differential equations for $\phi (z)$ and $b_{t} (z)$. We impose
the regular boundary condition at the horizon $z=1$. While near
the AdS boundary $z=0$, we can expand $\phi(z)$ and $b_t(z)$ as
\begin{equation}\label{eq:eps+7}
  \begin{aligned}
 \phi (z) & \approx  \phi _{s} z^{3-\Delta _{\phi }} +\phi _{o} z^{\Delta _{\phi}}+\cdots,  \\
b_{t}(z) & \approx  b_{s} z^{2-\Delta _{B}}   + b_{o}z^{\Delta _{B}-1}+\cdots,
  \end{aligned}
\end{equation}
where $\Delta _{\phi }=3/2 +\sqrt{9/4+m_{s }^2 l^2}$ and $\Delta
_{B }=2$. We turn off the source terms, namely
$\phi_s=b_s=0$. Whether there exist non-trivial
solutions to these equations depends on the wave number $k$ as
well as the Hawking temperature of black hole background. For
illustration, we take the mass of the dilaton
$m_{s}^{2}=-2/l^2=-8$. In Fig. \ref{fig1}, we plot the critical
temperature as the function of wave number. It is clearly seen
that the curve exhibits the bell curve behavior and the
unstable region looks like a dome. The highest critical
temperature is $T_{max}/ \mu \approx 0.0068 $, with wave number
$k_c/ \mu \approx 0.6856$. Within the dome, the AdS-RN black
hole becomes unstable and the charge density of gauge field $B$
becomes  spatially modulated.

\section{The instability of black holes with axion fields} \label{sec2}
\begin{figure} [h]
  \center{
  \includegraphics[scale=0.36]{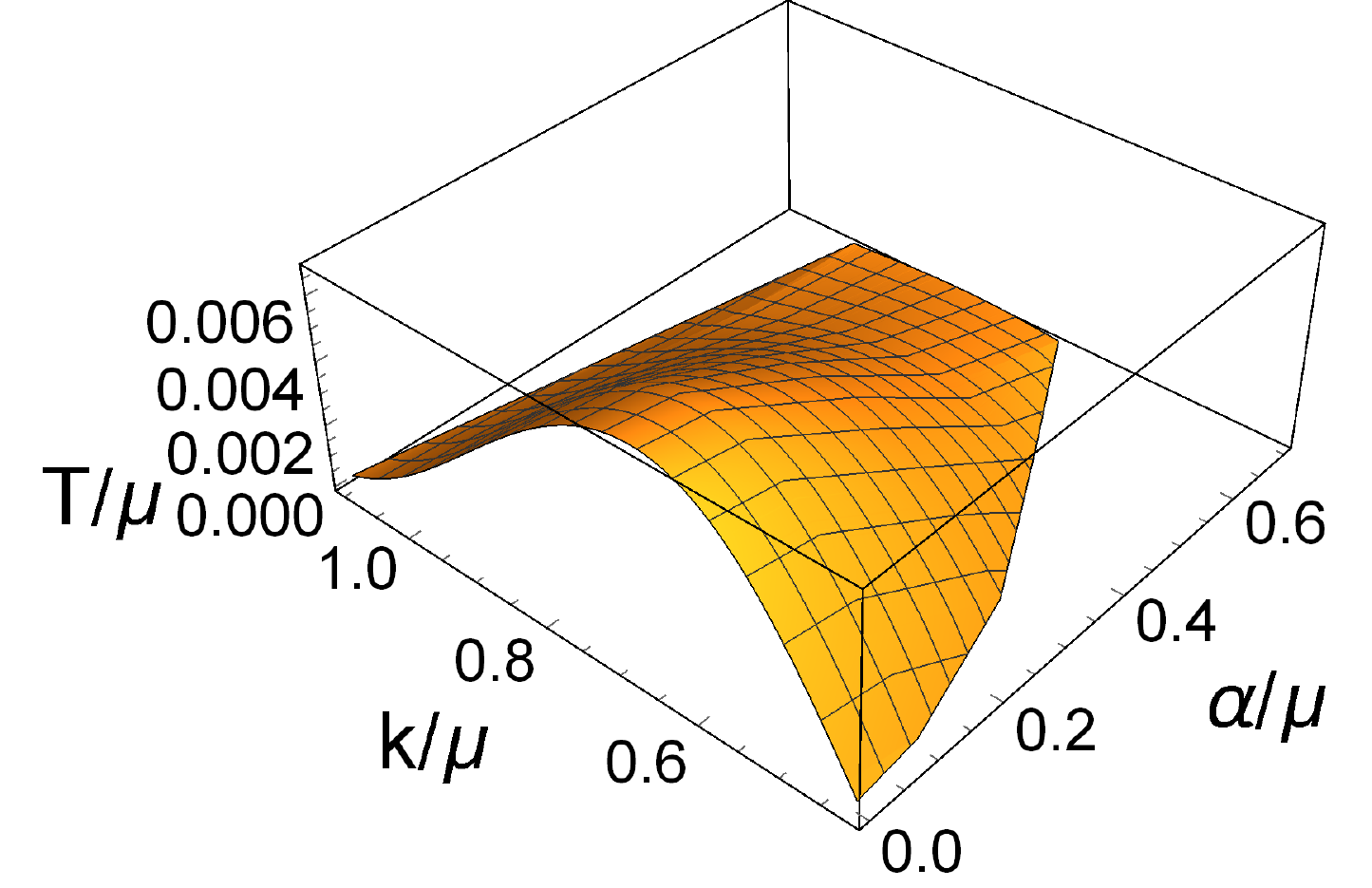}\ \hspace{0.1cm}
  \includegraphics[scale=0.28]{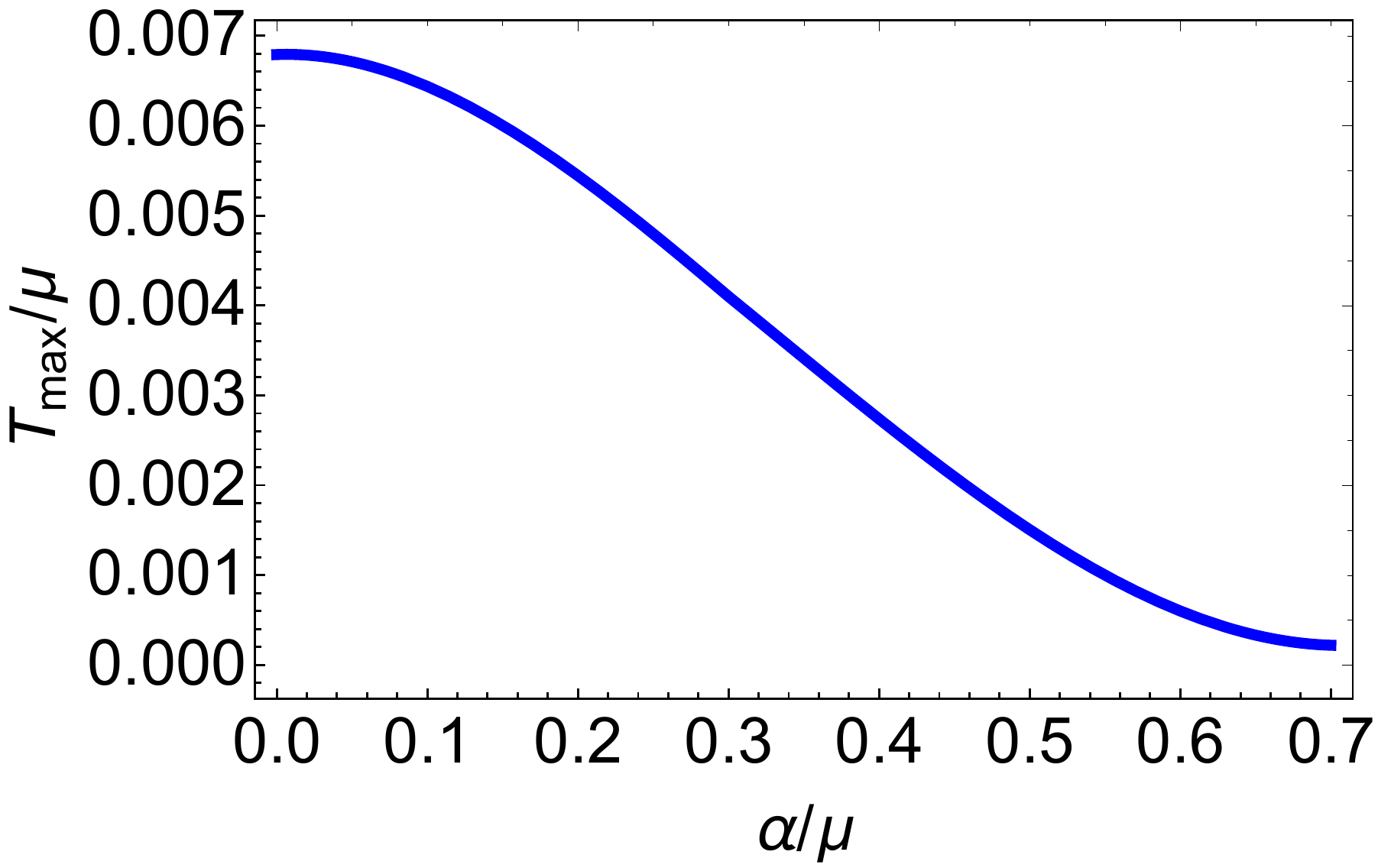}\ \hspace{0.1cm}
  \includegraphics[scale=0.27]{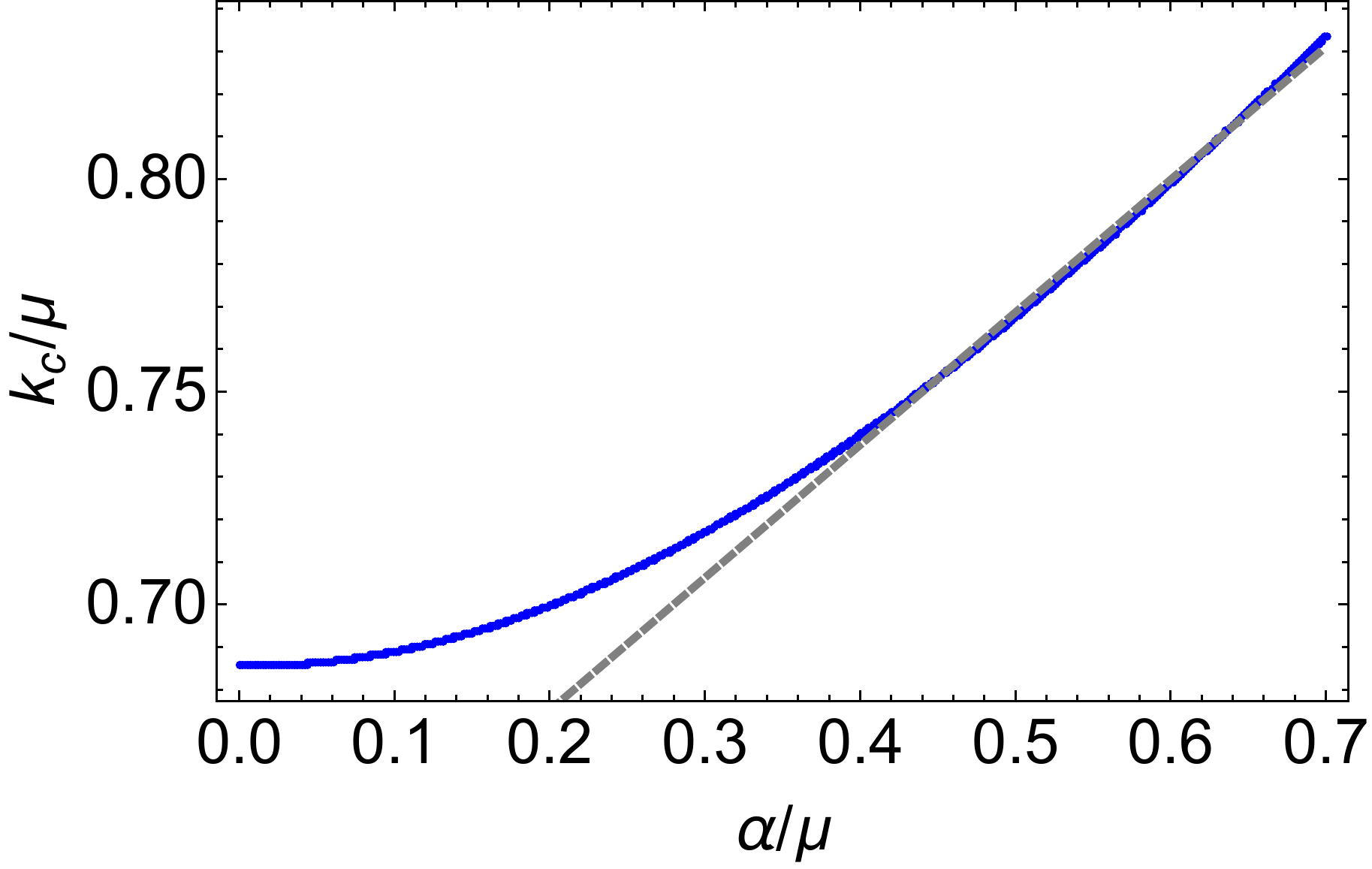}\ \hspace{0.1cm}
\caption{\label{fig3} The left 3D plot is the critical temperature
versus the wave number $k / \mu$ and the amplitude of the axion
field $\alpha/ \mu$. The middle plot is the highest critical
temperature $T_{max}/ \mu$ versus the amplitude of the axion field
$\alpha/ \mu$. The right plot is the wave number $k_c$ versus the
amplitude of the axion field $\alpha/ \mu$.
  }}
\end{figure}

In this section, we consider the instability of black holes
without translational invariance by adding axion fields into the
above action, which becomes
\begin{equation}\label{eq:eps+10}
  \begin{aligned}
  S=&\frac{1}{2\kappa^2}\int d^4 x \sqrt{-g} \ (\mathfrak{L_1}+\mathfrak{L}_{axions}),
  \end{aligned}
\end{equation}
with $\mathfrak{L_1}$ is given by (\ref{eq:eps+2}), and
\begin{equation}\label{eq:eps+11}
  \begin{aligned}
  \mathfrak{L}_{axions}&=& -\frac{1}{2}\sum_{a=1}^{2}(\nabla \chi _{a}
  ),
  \end{aligned}
\end{equation}
where $\chi _{a}$ are two real, massless scalar fields. This model
allows a very simple but exact solution with momentum relaxation,
which is given as
\begin{equation}\label{eq:eps+6}
  \begin{aligned}
     & ds^{2}=\frac{1}{z^2}\left[-(1-z)U(z)dt^{2}+\frac{dz^{2}}{(1-z)U(z)}+dx^{2} +dy^{2}\right], \\
     & A_t=\mu(1-z), \chi _{1} = \alpha x\ , \chi _{2} = \alpha y \ , \Phi=0, \
     B=0,
  \end{aligned}
\end{equation}
where $U(z)=4+4z -\frac{1}{2}(\alpha^{2} -8)z^{2} -\frac{1}{4}
\mu^{2} z^{3}$. The presence of the axion fields will introduce
momentum relaxation in both $x$ and $y$ directions, leading to a
finite DC conductivity\cite{Andrade:2013gsa}. Here we are
concerned with its impact on the instability of the background.
Without loss of generality, we consider the perturbations with
spatially modulated modes in $x$ direction. In the left of
Fig.\ref{fig3}, we plot the critical temperature versus the wave
number $k/\mu$ and the amplitude of axion fields $\alpha/ \mu$. It
is obvious that with the increase of  the amplitude $\alpha/ \mu$,
the unstable region becomes smaller, indicating that the
instability of the black hole is suppressed by the presence of the
axion field. We also plot the highest critical temperature
$T_{max}/ \mu$ and the corresponding wave number  $k_c/ \mu$ as a function of
 $\alpha/ \mu$, as illustrated in the middle and right of Fig.\ref{fig3}.
We notice that $k_c/ \mu$ grows linearly with large $\alpha/ \mu$,
while  $T_{max} / \mu$ drops down quickly with $\alpha/ \mu$. This
phenomenon is similar to what observed in other holographic
models\cite{Andrade:2015iyf}.

\section{The instability of black holes with Q-lattice}\label{sec3}
In this section we consider the instability of black holes with
Q-lattice\cite{Donos:2013eha}. Now the action becomes
\begin{equation}\label{eq:eps+10}
  \begin{aligned}
  S=&\frac{1}{2\kappa^2}\int d^4 x \sqrt{-g} \ (\mathfrak{L_1}+\mathfrak{L}_{Q}),
  \end{aligned}
\end{equation}
with $\mathfrak{L_1}$ in (\ref{eq:eps+2}) and
\begin{equation}\label{eq:eps+11}
  \begin{aligned}
  \mathfrak{L}_{Q}&=& -(\left | \nabla \Psi   \right |^{2} + m_{q} \left | \Psi   \right |^{2}
  ),
  \end{aligned}
\end{equation}
where $\Psi$ is a complex scalar field. Then the Einstein
equations become
\begin{equation}\label{eq:eps+13}
  \begin{aligned}
     & R_{\mu \nu }-T_{\mu \nu }^{\Phi }-T_{\mu \nu }^{A }-T_{\mu \nu }^{B }-T_{\mu \nu }^{AB  } -T_{\mu \nu }^{\Psi} =0,
  \end{aligned}
\end{equation}
with $T_{\mu \nu }^{\Phi }$ $T_{\mu \nu }^{A }$, $T_{\mu \nu }^{B }$, $T_{\mu \nu }^{AB }$ in (\ref{eq:eps+4}) and
\begin{equation}\label{eq:eps+14}
  \begin{aligned}
     & T_{\mu \nu }^{\Psi} =  \nabla_{\mu }\Psi  \nabla_{\nu }\Psi ^{*} + \frac{1}{2} m_{q}^{2}\left | \Psi  \right |^{2} g_{\mu \nu
     }.
  \end{aligned}
\end{equation}
In addition, we have an equation of motion for $\Psi$,
\begin{equation}\label{eq:eps+12}
  \begin{aligned}
 ( \nabla  ^{2} - m_{q})\Psi=0.
  \end{aligned}
\end{equation}
We consider the following ansatz for the electrically charged
AdS-RN black hole on Q-lattice,
\begin{equation}\label{eq:eps+15}
  \begin{aligned}
     & ds^{2}=\frac{1}{z^2}\left[-(1-z)p(z)Udt^{2}+\frac{dz^{2}}{(1-z)p(z)U}+ V_1 dx^{2} + V_2 dy^{2}\right], \\
     & A_t=\mu(1-z)a, \Psi=e^{ik_q x} z^{3-\Delta_q }\psi  , \ \Phi=0, \ B=0,
  \end{aligned}
\end{equation}
with $\Delta_q=3/2 \pm(9/4 + m_{q}^2 l^2)$. We choose the mass of
scalar field to be $m_{q}^{2}=-8$. Note that $U$, $V_1$,
$V_2$, $\psi$ and $a$ are function of the radial coordinate $z$
only. If one sets $U=V_1=V_2=a=1$ and $\psi=0$, then it goes back
to the standard AdS-RN black hole. For the non-trivial Q-lattice
solution, the boundary conditions at $z=0$ are given by
\begin{equation}\label{eq:eps+16}
  \begin{aligned}
U=V_1=V_2=a=1, \ \psi=\lambda  , \ \Phi=0, \ B=0,
  \end{aligned}
\end{equation}
and we take regular boundary condition on the horizon. As a
result, a typical black hole solution with Q-lattice is
characterized by three scale invariant parameters which are $T/
\mu$, $\lambda/ \mu$ and $k_q/ \mu$.
\begin{figure} [t]
  \center{
  \includegraphics[scale=0.6]{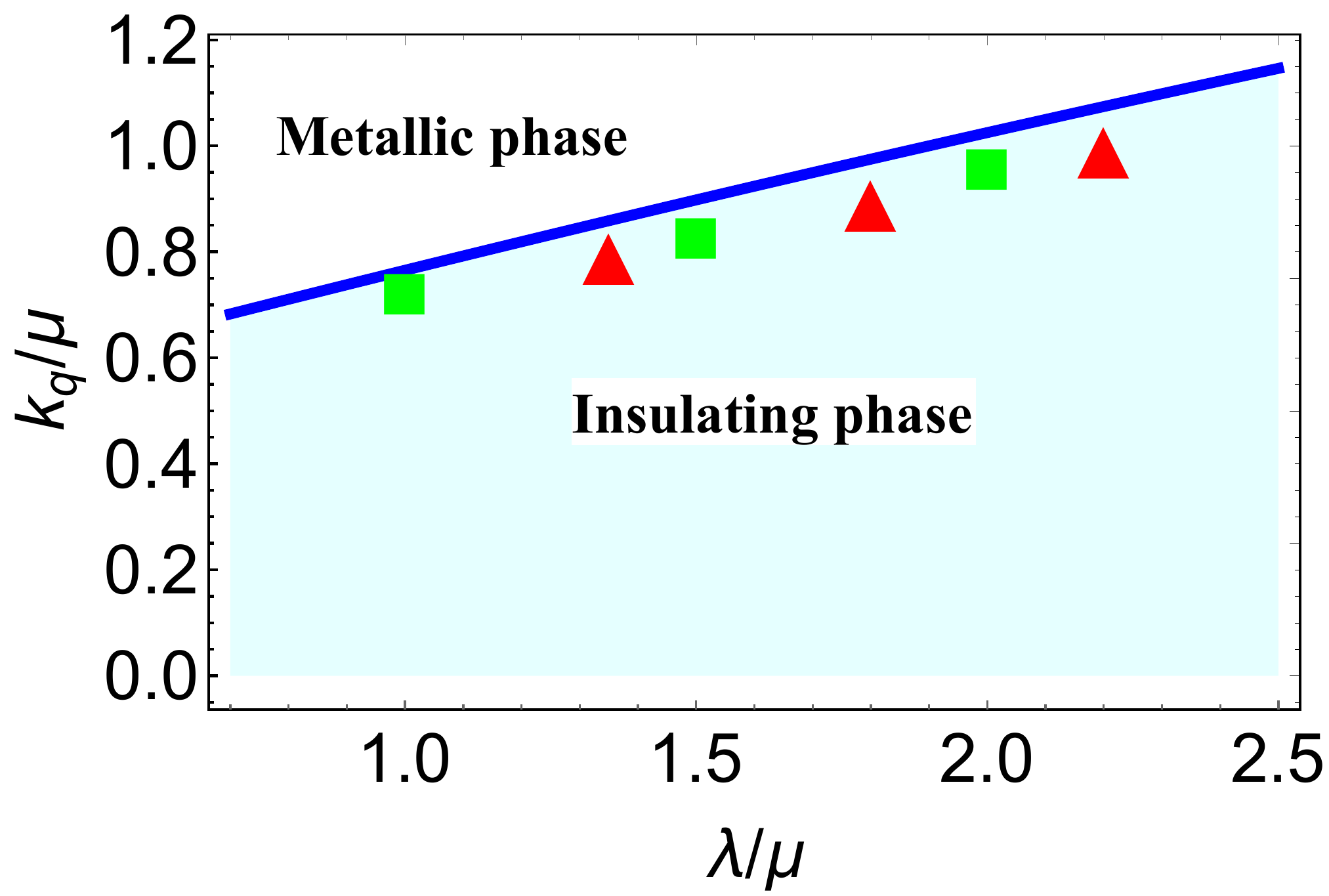}\ \hspace{0.1cm}
\caption{\label{fig5} The phase diagram for Q-lattice system
at temperature $T/\mu=0.001$. The red triangles stand for the
valleys of the highest critical temperatures $T_{max}/\mu$ when
varying $\lambda/ \mu$ but fixing $k_q/ \mu$. The green squares
stand for the valleys of the highest critical temperatures
$T_{max} /\mu$ when varying $k_q/ \mu$ but fixing $\lambda/ \mu$.
  }}
  \end{figure}

One remarkable feature of these Q-lattice background is that they
exhibit a novel metal/insulatore transition when we adjust the
lattice parameters $\lambda/ \mu$ and $k_q/ \mu$ in low
temperature limit\cite{Donos:2012js}, and the phase diagram for
metal/insulator phases over Q-lattice background has been studied
in Ref.\cite{Ling:2014laa,Ling:2015dma}. For a given background,
the DC conductivity can be expressed in terms of the horizon data
as
\begin{equation}\label{eq:eps+17}
  \begin{aligned}
\sigma _{DC} =(\sqrt{\frac{V_{2}}{V_{1}}}+\frac{\mu
^{2}a^{2}\sqrt{V_{1}V_{2}}}{k_q^{2} \psi^{2} } )|_{z=1}.
  \end{aligned}
\end{equation}
Throughout this paper we identify the metallic phase and the
insulating phase by evaluating Eq.(\ref{eq:eps+17}) around
$T/\mu=0.001$. The quantum critical line is given by $\partial
_{T} \sigma _{DC} = 0$. In Fig. \ref{fig5}, we demonstrate the
phase diagram for Q-lattice system.
\begin{figure} [t]
  \center{
  \includegraphics[scale=0.22]{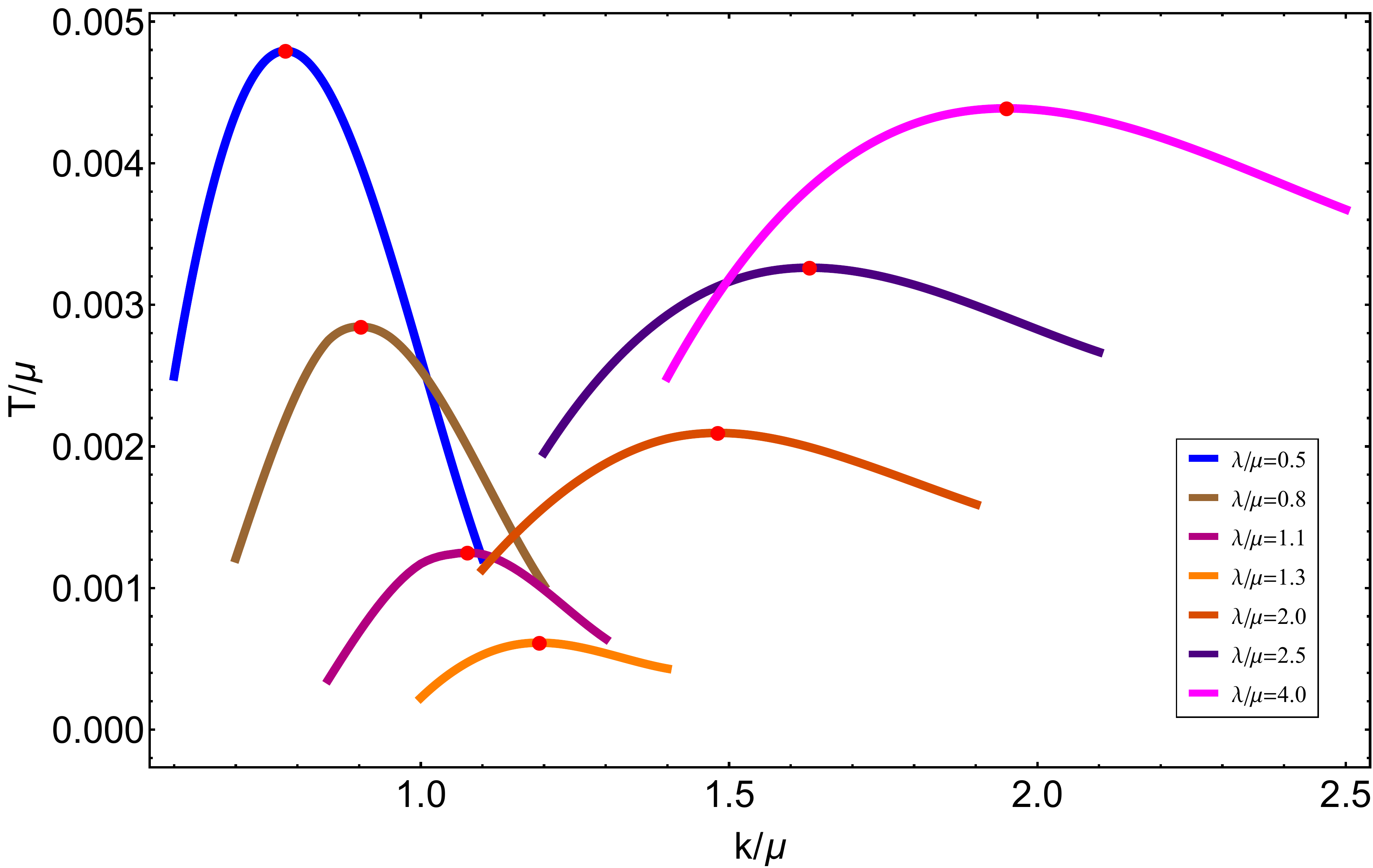}\ \hspace{0.1cm}
  \includegraphics[scale=0.22]{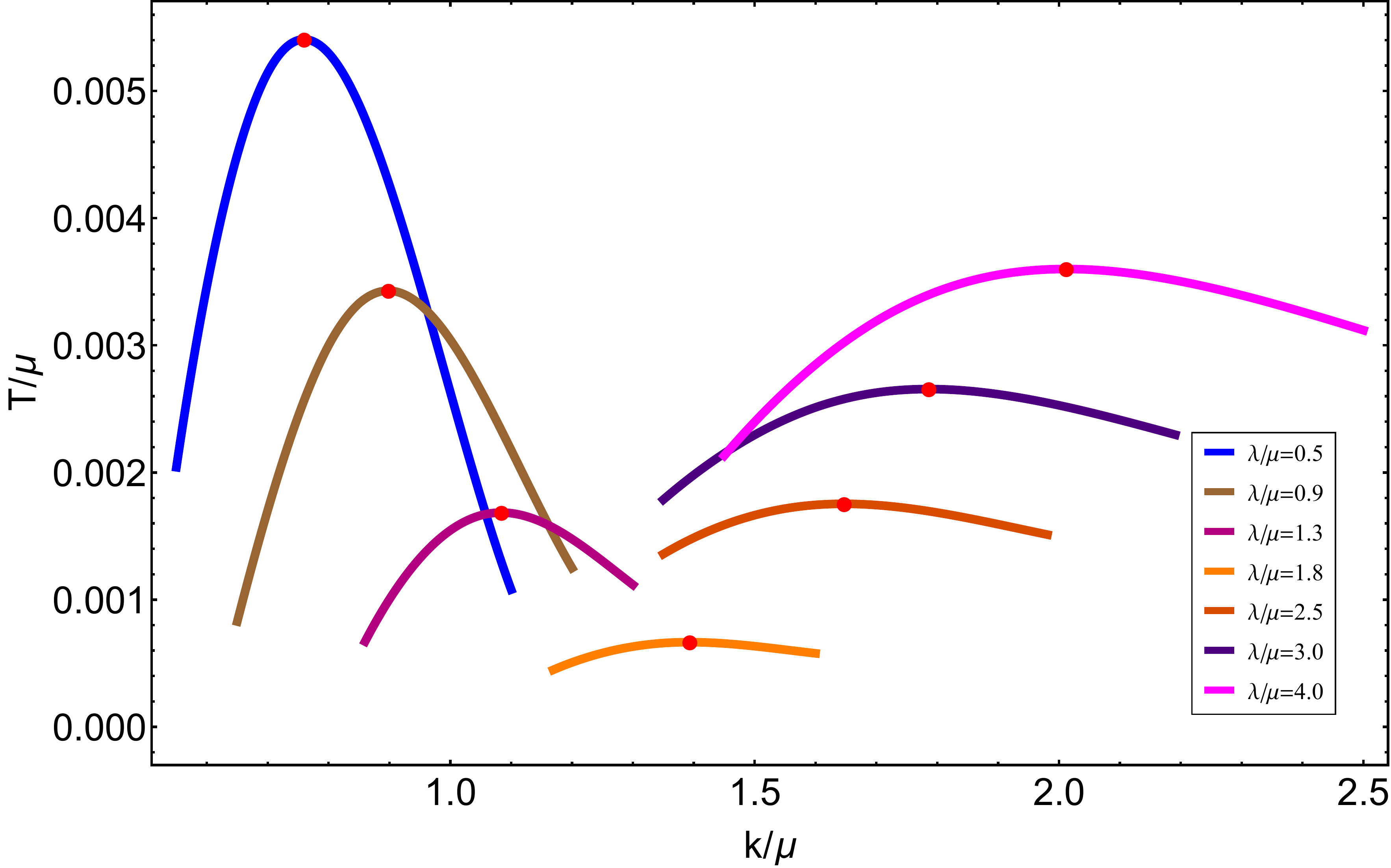}\ \hspace{0.1cm}
\caption{\label{fig2} The unstable region of the background with
different values of $\lambda/ \mu$ when the wave number of
Q-lattice $k_q$ is fixed as $k_q/ \mu=0.8$(left) and
$k_q/\mu=0.9$(right). The red dots mark the highest critical
temperature on each curve with the corresponding wave number
$k_c/ \mu$.
  }}
  \end{figure}

\begin{figure} [t]
  \center{
  \includegraphics[scale=0.38]{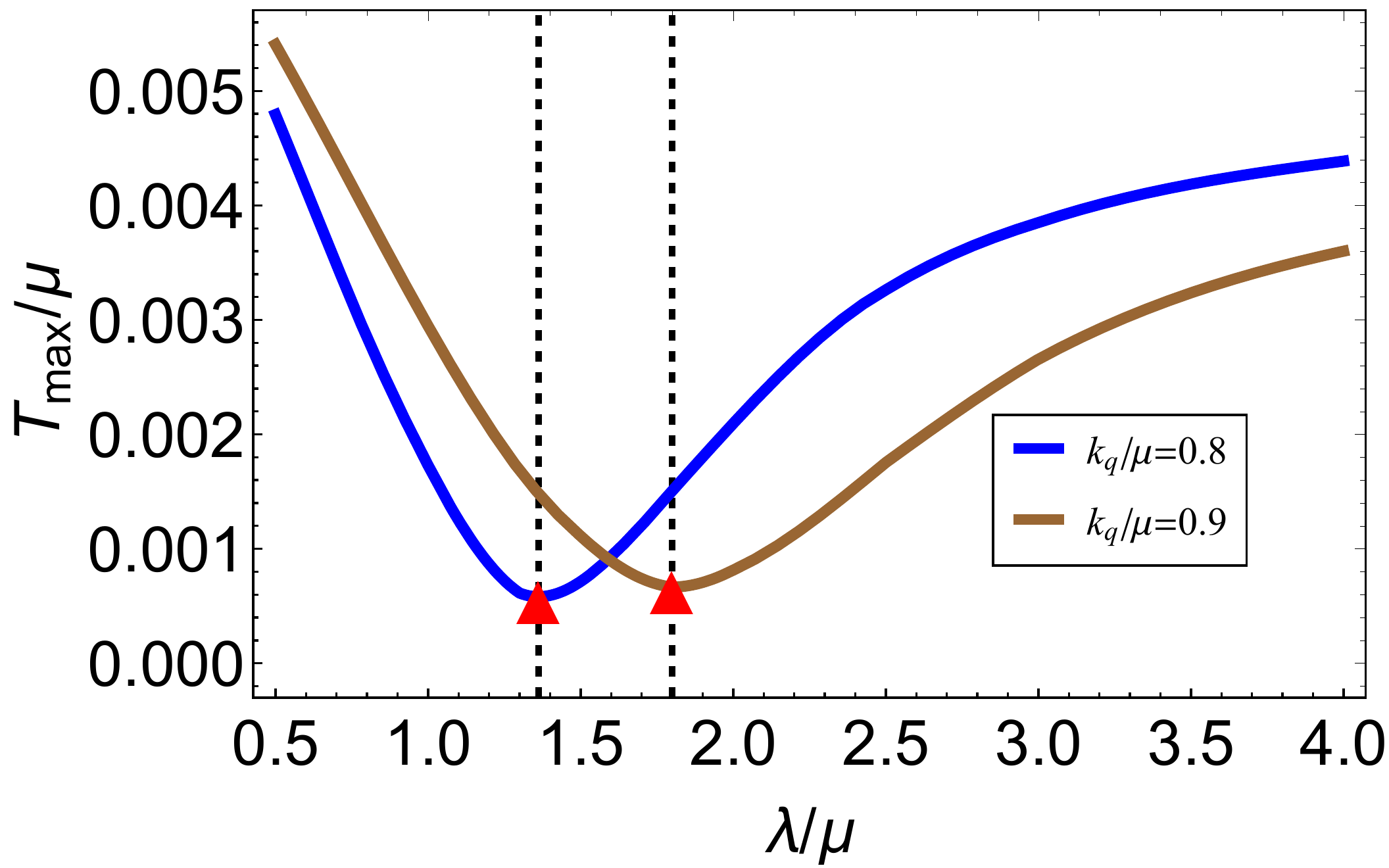}\ \hspace{0.1cm}
  \includegraphics[scale=0.37]{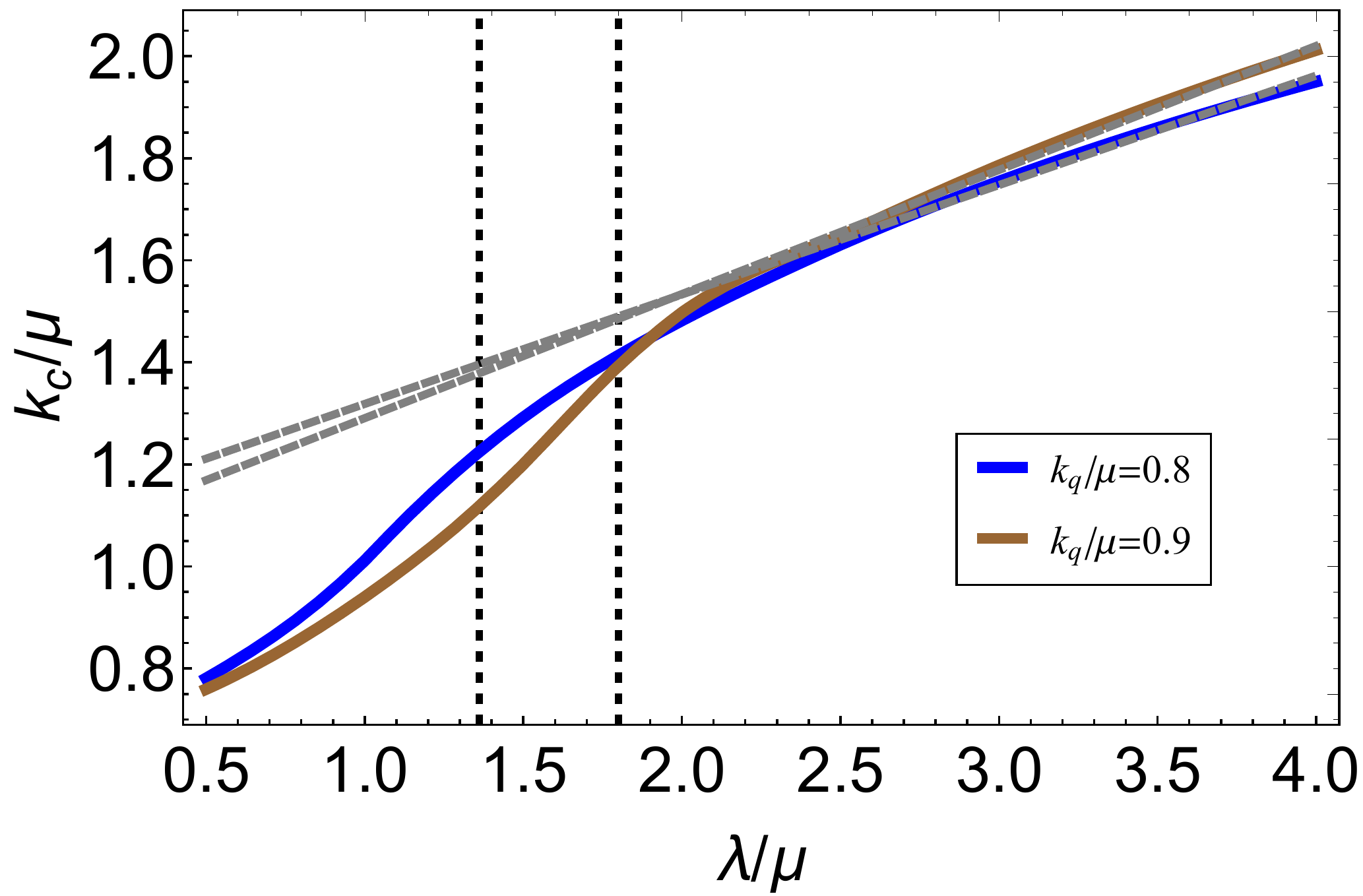}\ \hspace{0.1cm}
\caption{\label{fig6} The highest critical temperature
$T_{max}/\mu$ and the wave number $k_c/\mu$ as the function of the
lattice amplitude $\lambda/ \mu$ for the given wave number $k_q/
\mu=0.8$ and $k_q/\mu=0.9$.
  }}
  \end{figure}

\begin{figure} [h]
  \center{
  \includegraphics[scale=0.22]{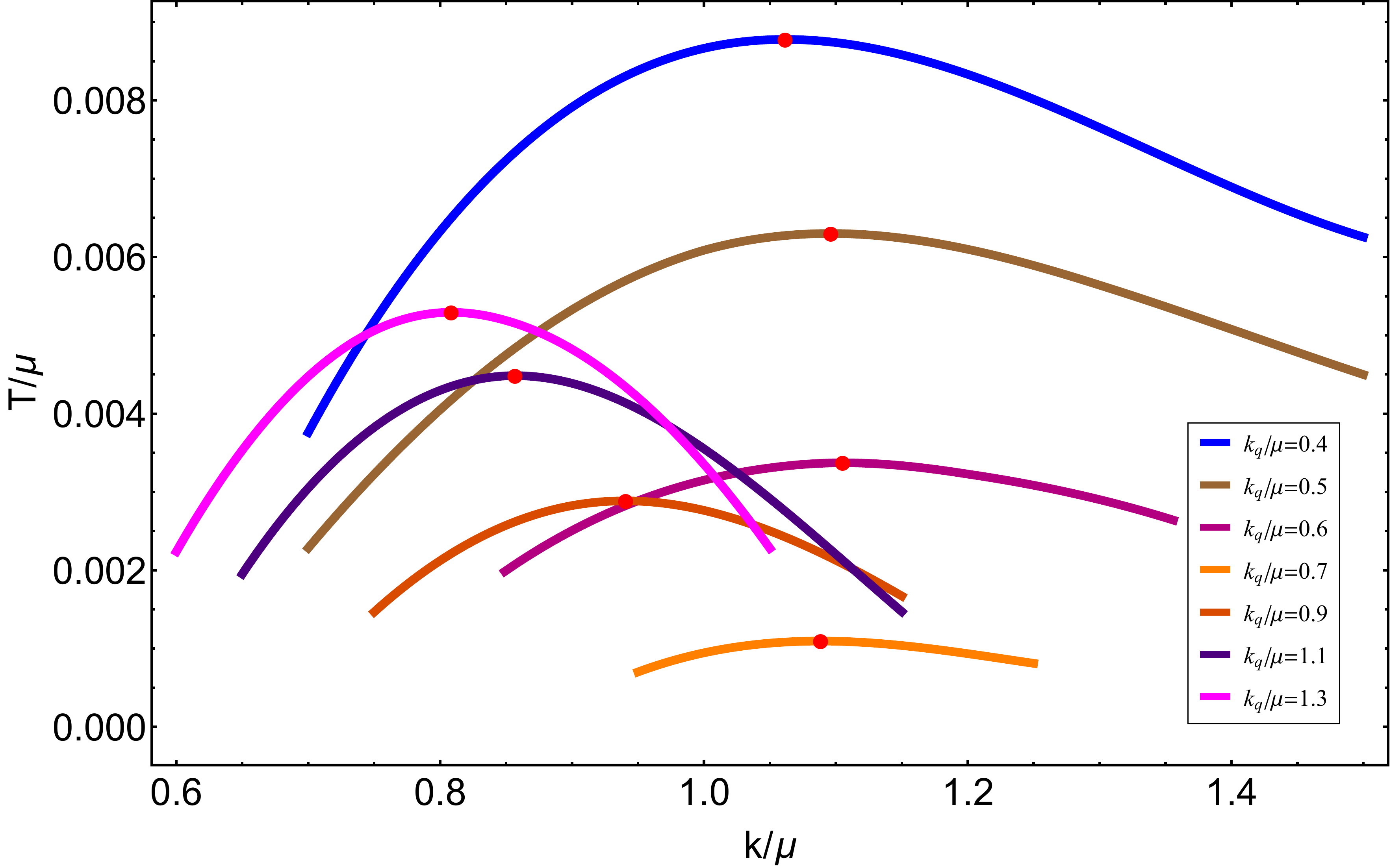}\ \hspace{0.1cm}
  \includegraphics[scale=0.22]{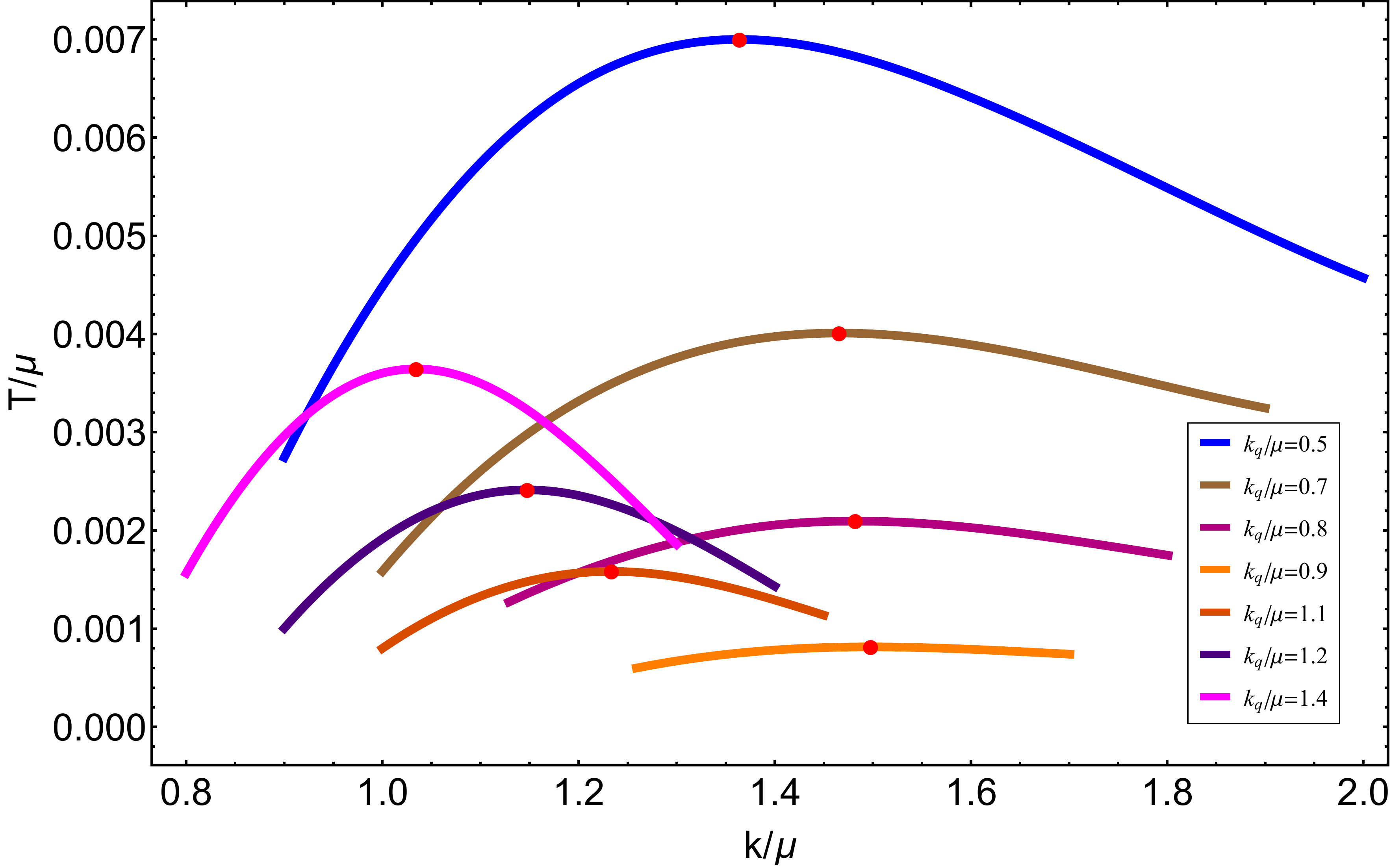}\ \hspace{0.1cm}
\caption{\label{fig7} The unstable region of the background with
different values of $k_q/ \mu$ when the lattice amplitude
$\lambda/ \mu$ is fixed as $\lambda/ \mu=1$(left) and
$\lambda/\mu=2$(right). The red dots mark the highest
critical temperature on each curve with the corresponding
wave number $k_c/ \mu$.
  }}
  \end{figure}

\begin{figure} [h]
  \center{
  \includegraphics[scale=0.38]{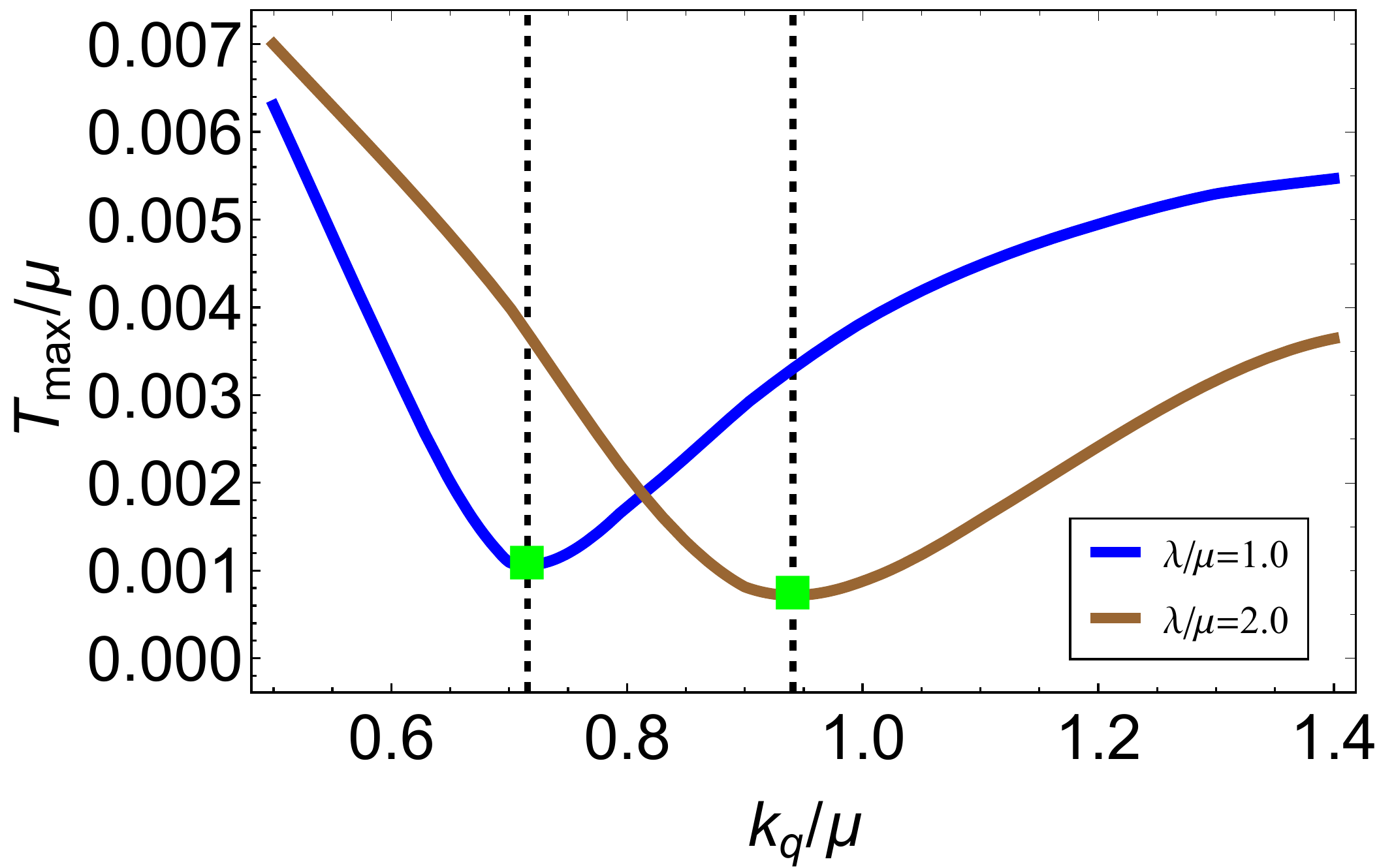}\ \hspace{0.1cm}
  \includegraphics[scale=0.37]{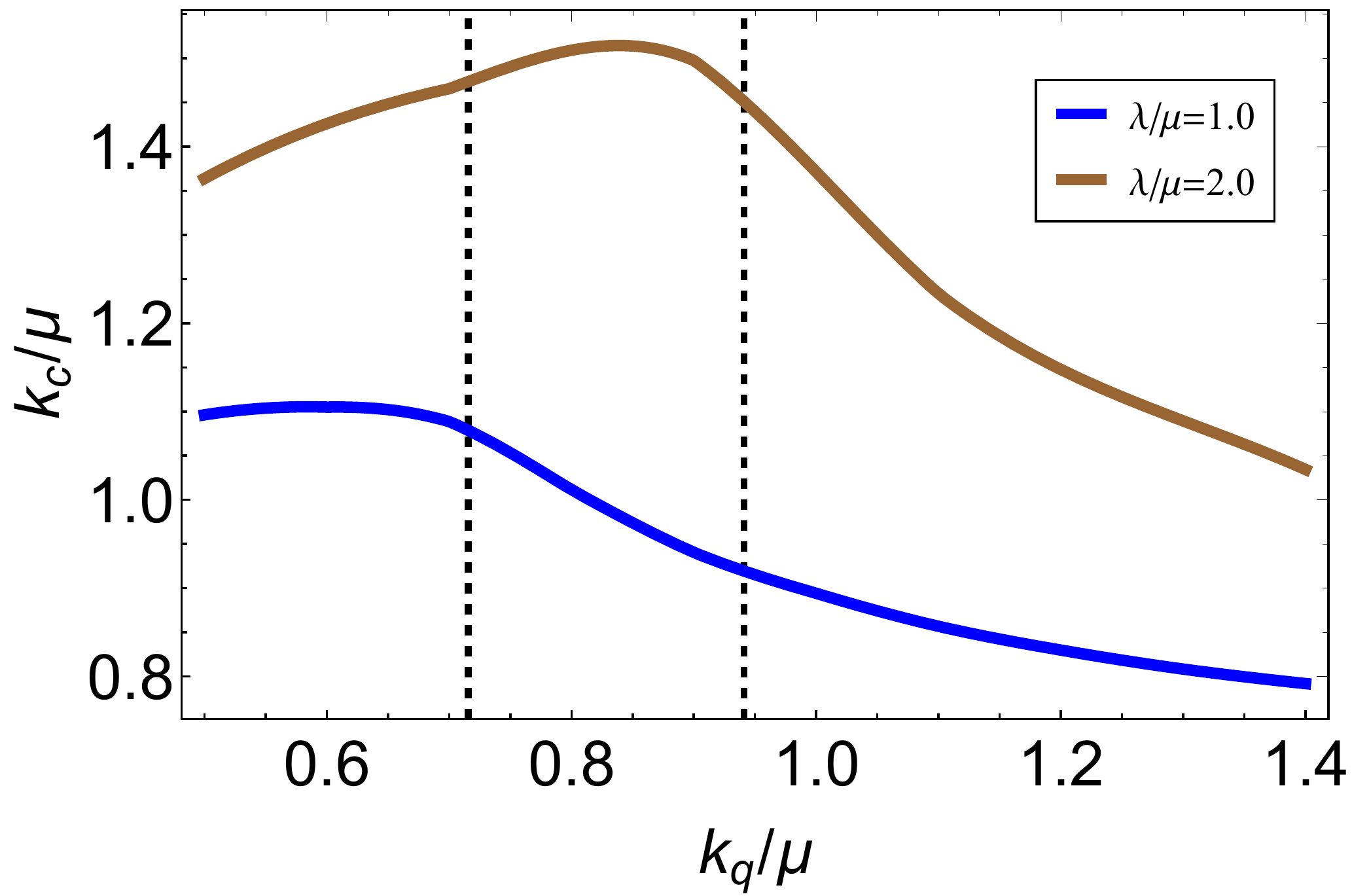}\ \hspace{0.1cm}
\caption{\label{fig8} The highest critical temperature
$T_{max}/\mu$ and the  wave number $k_c/\mu$ as the function of
the lattice wave number $k_q/ \mu$ for the given amplitude
$\lambda/ \mu=1$ and $\lambda/ \mu=2$.
  }}
  \end{figure}

Now we focus on the instability of this sort of black holes with
lattices. In Fig. \ref{fig2}, we plot the unstable region of the
background with different values of $\lambda/ \mu$ when the wave
number of Q-lattice $k_q/\mu$ is fixed as $k_q/ \mu=0.8$ and
$k_q/\mu=0.9$, respectively. The region below each curve is the
unstable region where the spatially modulated modes with wave
number $k/\mu$ may emerge. In contrast to the results observed in
axion model in previous section, we find the unstable region does
not change monotonously with the lattice parameter any more. To
our surprise, we find the unstable region becomes smaller with the
increase of $\lambda/\mu$ at first, but later it becomes larger
again. To disclose this with more transparency, we may mark the
highest critical temperature $T_{max}/\mu$ with red dot on
each curve and denote the corresponding wave number as $k_c/\mu$.
Then we plot $T_{max}/\mu$ and $k_c/\mu$ as the function of
the lattice amplitude $\lambda/ \mu$, as illustrated in Fig.
\ref{fig6}. It is obvious to see that $T_{max}/\mu$ reaches
the minimal value and then rises up again with the increase of
$\lambda/ \mu$. Going back to the phase diagram of the Q-lattice,
we find the turning points are quite close to the critical line
for metal/insulator transition! To check this we may also plot the
unstable region of the background with different values of $k_q/
\mu$ but fixing the lattice amplitude $\lambda/ \mu$, as
illustrated in Fig.\ref{fig7}. Correspondingly,
$T_{max}/\mu$ and $k_c/\mu$ as the function of the wave number
$k_q/ \mu$ is plotted in Fig.\ref{fig8}. Again, we find that the
turning points of  $T_{max}/\mu$ are quite close to the
critical line, as marked in Fig.\ref{fig5}. Since the curve with
the minimal value of  $T_{max}/\mu$ encloses the smallest
region of instability, as illustrated in Fig. \ref{fig2}, we
conclude that the black hole background in the vicinity of
metal/insulator transition is the most stable solutions under the
perturbations of spatial modulated modes given by
(\ref{eq:eps+7}).

First of all, it is quite interesting to compare the change of
$T_{max}/ \mu$ of CDW with that of superconductivity with
the change of Q-lattice parameters, which was previously
investigated in \cite{Ling:2014laa}. It is found that the critical
temperature of superconductivity is always suppressed by the
presence of the Q-lattice. When the lattice effect is strong
enough, the critical temperature of superconductivity drops
down to zero such that the superconducting phase disappears.
However, for CDW the critical temperature will rise up again with
the increase of lattice parameters, which means the background
becomes more unstable and it is easier to form a new background
with CDWs. This difference may be understood based on the phase
diagram of Q-lattice background as demonstrated in
Fig.\ref{fig5}. For large $\lambda/\mu$, the dual system falls
into a deep insulating phase. Therefore, it is quite nature to
understand that it becomes harder to form superconductivity over
such insulating phases. On the contrary, the CDW phase itself is
an insulating phase. Thus such a background dual to a deep
insulating phase will assist the formation of CDW.

This phenomenon indicates that the instability of the background
might be used to characterize the occurrence of quantum phase
transition. It is well known that in the absence of ordinary order
parameters, it is very hard to diagnose quantum phase transitions.
Previously the role of holographic entanglement entropy in
diagnosing the quantum phase transition has been disclosed in a
series of
papers\cite{Ling:2015dma,Ling:2016wyr,Ling:2016dck}. The
holographic entanglement entropy or its derivative with respect to
system parameters displays a peak or valley in the vicinity of the
critical region. Here we find $T_{max}/ \mu$ of CDW exhibits
a similar behavior as the entanglement entropy. Physically, it
implies that the system becomes rather stable under the
perturbations with spatially modulated modes near the critical
point of quantum phase transition in zero temperature limit.
Intuitively, this might be understood as follows. The formation of
CDW over a fixed background results from a thermodynamical phase
transition. Near the critical point of quantum phase transition,
the system becomes long-range correlated such that the effects of
thermal perturbations would be suppressed, leading to a more
stable background with the lowest $T_{max}/ \mu$.

Finally, we may also give a comment on the commensurability of
this sort of black holes with lattice, which involves in the
comparison between the wave number of CDW, namely $k_c/ \mu$ and
the wave number of lattices. We plot $k_c/ \mu$ as the function
of the lattice amplitude $\lambda/ \mu$ in Fig. \ref{fig6} and the
wave number of Q-lattice $k_q/ \mu$ in Fig. \ref{fig8},
respectively. We find that $k_c/ \mu$ grows linearly with the
amplitude of Q-lattice $\lambda/ \mu$ for large $\lambda/ \mu$.
However, it decreases with the wave number of Q-lattice $k_q/ \mu$
for large $k_q/ \mu$. No manifest effect or phenomenon is observed
when $k_c/ \mu=k_q/ \mu$. The commensurability seems absent in
this setup, similar to the result observed in
\cite{Andrade:2015iyf}.

\section{Discussion}\label{sec4}
In this paper we have investigated the instability of black holes
with momentum relaxation. It is found that the presence of linear
axion fields suppresses the instability of AdS-RN black hole. The
wave number $k_c/  \mu$ of spatially modulated modes grows
linearly with the axion field $\alpha/ \mu$ for large $\alpha/
\mu$. More importantly, in Q-lattice framework we have
demonstrated that in zero temperature limit the unstable
dome is the smallest near the critical region of
metal/insulator transition. The highest critical temperature
$T_{max}/\mu$ displays a valley near the critical
points of metal/insulator transition. This novel phenomenon
is reminiscent of the behavior of the holographic entanglement
entropy during quantum phase transition. We conjecture that any
instability of background leading to a thermodynamical phase
transition would be greatly suppressed in the critical region of
quantum phase transition, since in this region the system becomes
long-range correlated. The commensurate effect is absent in both
models with homogeneous lattices, similar to the results
obtained in Ref.\cite{Andrade:2015iyf}.

In this paper we have only presented the perturbation
analysis over a fixed background to justify the instability of
this background. To explicitly construct a new background with
both lattices and CDW, one needs to go beyond the
perturbation analysis and solve all the equations of motion
numerically, which are PDEs rather than ODEs. It is completely
plausible to obtain such solutions at normal temperatures, as
investigated in \cite{Donos:2013gda,Ling:2014saa}. However, as we
mentioned in the introduction, finding numerical solutions would
become rather difficult in zero temperature limit. Our analysis on
the instability of black holes in this paper sheds light on the
construction of CDW background with Q-lattice in zero temperature
limit since the critical temperature and the unstable region have
been manifestly disclosed at extremely low temperature.

The phase diagram for high $T_c$ superconductivity exhibits a
very abundant structure with many universal features. Currently it
is still challenging to exactly duplicate this phase diagram in
holographic approach. Based on our current work, one may further
introduce the complex scalar field as the order parameter of
superconductivity, and consider the condensation of
superconductivity due to $U(1)$ gauge symmetry breaking. Then it
would be quite interesting to investigate the relations between
CDW and superconductivity over such a lattice background.

\centerline{\rule{80mm}{0.1pt}}

\end{document}